\documentclass[a4paper]{article}

\usepackage{INTERSPEECH2021}
\usepackage{comment}

\def\x{{\mathbf x}}
\def\y{{\mathbf y}}
\def\s{{\mathbf s}}
\def\z{{\mathbf z}}
\def\g{{\mathbf g}}
\def\X{{\mathbf X}}
\def\S{{\mathbf S}}
\newcommand{\argmax}{\mathop{\rm arg~max}\limits}

\title{Data Augmentation Methods for \\End-to-end Speech Recognition on Distant-Talk Scenarios}
\name{Emiru Tsunoo$^1$, Kentaro Shibata$^1$, Chaitanya Narisetty$^2$, Yosuke Kashiwagi$^1$, Shinji Watanabe$^2$}
\address{
  $^1$Sony Corporation\\
  $^2$Carnegie Mellon University}
\email{Emiru.Tsunoo@sony.com}

\begin{document}

\maketitle
\begin{abstract}
Although end-to-end automatic speech recognition (E2E ASR) has achieved great performance in tasks that have numerous paired data, it is still challenging to make E2E ASR robust against noisy and low-resource conditions.
In this study, we investigated data augmentation methods for E2E ASR in distant-talk scenarios.
%A recurrent neural network transducer (RNN-T) is trained on the CHiME-6 Challenge dataset, which is a suitable task for studying robustness against noisy and spontaneous speech.
E2E ASR models are trained on the series of CHiME challenge datasets, which are suitable tasks for studying robustness against noisy and spontaneous speech.
We propose to use three augmentation methods and thier combinations: 1) data augmentation using text-to-speech (TTS) data, 2) cycle-consistent generative adversarial network (Cycle-GAN) augmentation trained to map two different audio characteristics, the one of clean speech and of noisy recordings, to match the testing condition, and 3) pseudo-label augmentation provided by the pretrained ASR module for smoothing label distributions.
Experimental results using the CHiME-6/CHiME-4 datasets show that each augmentation method individually improves the accuracy on top of the conventional SpecAugment; further improvements are obtained by combining these approaches.
We achieved 4.3\% word error rate (WER) reduction, which was more significant than that of the SpecAugment, when we combine all three augmentations for the CHiME-6 task.

\end{abstract}

\section{Introduction}
\label{sec:intro}
End-to-end automatic speech recognition (E2E ASR), which is an integrated neural network that directly estimates output text sequences from audio features, has attracted considerable attention.
E2E ASR systems have many variations, such as connectionist temporal classification (CTC) \cite{graves06%, miao15%,amodei16
}, attention-based encoder--decoder models \cite{chorowski15, chan16%, chiu18
}, hybrid models \cite{watanabe17}, recurrent neural network transducers (RNN-Ts) \cite{graves13rnnt%,rao17
}, Transformers \cite{%vaswani17,
karita19}, and Conformers \cite{gulati2020}.
These E2E approaches achieve excellent performance, particularly when they exploit a large amount of training data.

Given such insights, data augmentation is known to be effective for E2E ASR. %, which is a many-to-one problem.
Conventionally, speed perturbation \cite{ko2015audio} and vocal tract length perturbation (VLTP) \cite{jaitly2013vocal} are widely used.
SpecAugmentation \cite{park19}, which applies on-the-fly time warping and frequency/time masking, constantly achieves significant improvements in various tasks.
%Because the text-to-speech (TTS) task is a one-to-many problem, 
Because the text-to-speech (TTS) generates various speech styles, it is suitable for augmentation, and the synthesized data is used in \cite{hayashi2018back,tjandra2019end, wang2020improving%, rossenbach2020generating
}.
% Further, SCADA \cite{wang2020scada} includes virtual adversarial training to increase data variation.
For semi-supervised learning, many label augmentation techniques have been proposed, where pseudo labels are generated by a model trained with limited supervised data \cite{kemp1999unsupervised,%,parthasarathi2019lessons,
chen2020semi}. %,xu2020iterative}.
%To recognize whisper speeches, Guedepu {\it et al.} augmented training data with a synthetic whisper corpus generated from normal speech using a cycle-consistent generative adversarial network (Cycle-GAN) \cite{gudepu2020whisper}.

However, in the case of distant spontaneous talk in noisy environments with low resources, the situation becomes severe.
Based on the solutions to the CHiME-6 Challenge, which is a competition to solve the aforementioned problem, only conventional hybrid systems have made a meaningful attempt in this regard \cite{du2020ustc,chen2020ioa%,medennikov2020stc
}.
Follow-up studies have investigated E2E approaches on tasks but have failed to outperform the hybrid systems \cite{Dalmia2018,yalta2019cnn}.
Recently Andrusenko {\it et al.} explored model architectures; it was concluded that RNN-T achieved comparable performance to hybrid approaches \cite{andrusenko2020towards}.
Although many augmentation methods have been applied to the task, there is still scope for further exploration.

In general, there are not enough in-domain conversational data because conversational styles have many variations in nature.
Therefore, augmentation using TTS-synthesized speeches is suitable to help this regard.
Further, in many cases, simulating noise environments does not match the real testing environments.
Cycle-consistent generative adversarial network (Cycle-GAN) \cite{zhu2017unpaired} is a way to map two different domains; thus this can be used to transform from clean audio characteristics to the ones in noisy environments.
Conversational distant-talk speech also has a difficulty in transcription such as label errors and recording failures.
Therefore, the reference label distribution is biased and does not represent conversational speech properly.
Pseudo labels generated by a statistical model, such as pretrained ASR, mitigate such label errors and biases.

In this study, we investigated data augmentation methods for E2E ASR in distant-talk scenarios.
% RNN-T is trained on the CHiME-6 Challenge dataset to study its robustness against noisy and spontaneous speech.
% The latest Conformer architecture is also trained on the CHiME-4 Challenge dataset.
CHiME-6/CHiME-4 Challenge datasets are used to study its robustness against noisy and spontaneous speech.
We propose to use aforementioned three augmentation methods and their combinations.
%: data augmentation using synthesized TTS speeches, Cycle-GAN trained to map from binaural clean recordings to multi-microphone noisy recordings to match the testing condition, and pseudo-label augmentation provided by the pretrained ASR model for label distribution smoothing.
Our contributions are as follows.  
1)~This study is the first to apply TTS augmentation to distant-talk problems like CHiME challenges.
2)~It is also the first study to use Cycle-GAN to map training data from clean audio to noisy distant-talk for augmentation.
3)~We investigate the usage of pseudo labels in supervised learning scenarios for data augmentation.
4)~The combination of augmentation methods on top of the SpecAugment achieved 4.3\% WER reduction from only the SpecAugment in the CHiME-6 task and 1.3/0.5\% reduction in the CHiME-4 simulation/real tasks.

\begin{figure*}[t]
%\vspace{-3.5cm}
  \centering
  %\hspace{1.5cm}
  \includegraphics[width=0.72\textwidth]{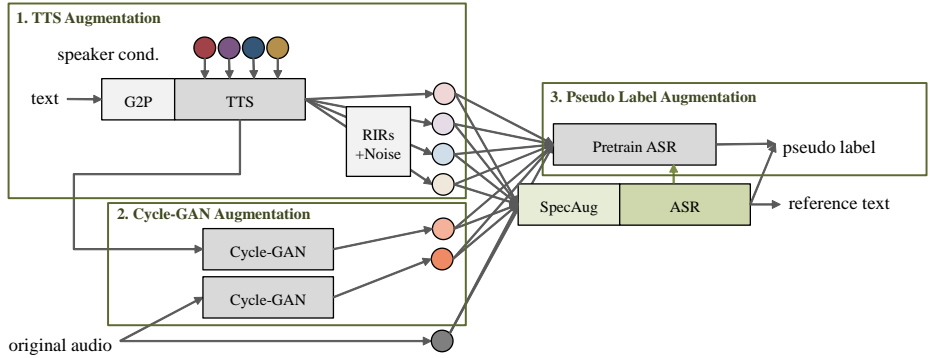}
  \vspace{-0.3cm}
  \caption{Overview of augmentation methods for distant-talk ASR.}
  \label{fig:augmentation}
  \vspace{-0.3cm}
\end{figure*}

\section{Distant-Talk Scenarios}
In the distant-talk scenarios, where speech is heavily contaminated by reverberation and noises, it is more challenging for E2E ASR to acquire robustness with only limited audio resources.
The series of CHiME Challenge is suitable for investigating in this regard; thus we use CHiME-6/CHiME-4 Challenges in this study.

\subsection{CHiME-6 Challenge}
The CHiME-6 Challenge \cite{watanabe2020chime} targets the problem of distant conversational ASR in everyday home environments. 
Conversations in twenty real dinner parties were recorded with multiple Microsoft Kinect having 4-channel microphone arrays and were fully transcribed.
In addition, binaural microphones were worn by the participants.
%These recordings can only be used in the training stage.
%In the training step, binaural recordings are perturbed with various room impulse responses (RIRs) generated by a room simulator and additive noise, and subsequently combined with Kinect recordings as the training set.
The close-talk binaural recordings play an important role for building an acoustic model as it is cleaner than the Kinect recordings.
In the evaluation step, guided source separation (GSS) based speech enhancement preprocess is applied \cite{boeddeker2018front}.
%Although the task is challenging to E2E approaches, Andrusenko {\it et al.} achieved comparable performance with E2E RNN-T architecture \cite{andrusenko2020towards}.

\begin{comment}
The baseline system is provided using the following setups:
First, microphone arrays are synchronized to remove frame drops and clock-drifting.
Binaural recordings are perturbed with various room impulse responses (RIRs) generated by a room simulator and additive noise, and subsequently combined with Kinect recordings as the training set.
Triphone GMM-HMM models were trained and used to cleanup the training set.
Subsequently, the factorized time-delayed neural network (TDNN-F) ASR \cite{povey2018semi} is trained with lattice-free maximum mutual information (LF-MMI) criterion \cite{povey16}.
In the evaluation step, guided source separation (GSS) based speech enhancement preprocess is applied \cite{boeddeker2018front}.
Although the task is challenging to E2E approaches, Andrusenko {\it et al.} achieved comparable performance with E2E RNN-T architecture \cite{andrusenko2020towards}.
\end{comment}

\subsection{CHiME-4 Challenge}
In the preceding CHiME-4 Challenge \cite{vincent2017analysis}, WSJ prompts were recorded by 6 microphones embedded in a tablet device with 4 noisy locations, i.e., on the bus, cafe, pedestrian area and street junction.
Along with the real recordings, simulation data is composed for 4 noisy environments by mixing with the original WSJ data or the booth recordings.
We use the CHiME-4 Challenge data to confirm that our proposed augmentation methods are effective for different tasks and model architectures.
Only isolated single-channel tracks are used for this purpose.

\section{Augmentation Methods}
For distant-talk low-resource scenarios, we propose to use three augmentation methods to train a robust E2E ASR model.
%First, data augmentation is performed using TTS data, which is a novel approach for such a distant-talk task. 
%Second, Cycle-GAN is also used as an additional augmentation to close the gap between the training audio characteristics and testing noisy characteristics. 
%Finally because reference labels are also noisy owing to spontaneous speech, pseudo label augmentation given by the pretrained ASR module is applied to smoothen the label distributions.
An overview of the augmentation methods is shown in Fig.~\ref{fig:augmentation}.

\subsection{Text-to-speech data augmentation}
\label{ssec:tts}
TTS synthesized data augmentation has recently gained popularity \cite{hayashi2018back, tjandra2019end, wang2020improving%, rossenbach2020generating
}.
It is a one-to-many mapping problem; thus it is suitable for augmenting data with a broad variation coverage.
We use a Transformer TTS \cite{li2019neural} conditioned on speaker-conditioning information, i.e., x-vector \cite{snyder2018x}, and a global-style token (GST) \cite{wang2018style}.
Let the target label sequence be $\y$, the x-vector be $\z$, and the GST be $\g$, then the paired data of the synthesized audio signal sequence $\hat{\s}^{\mathsf{tts}}$ and its label is obtained as
\begin{align}
    (\hat{\s}^{\mathsf{tts}},\y) \leftarrow \hat{\s}^{\mathsf{tts}} \sim %=\argmax_{\s}
    q_{\mathrm{tts}}(\s|\y,\z,\g),\label{eq:tts}
\end{align}
where $\z$ and $\g$ are added to the hidden output of the encoder in the Transformer.

Because the training dataset is noisy and small, it is unfeasible to train a TTS model.
Therefore, we use a TTS model trained with external Librispeech dataset \cite{panayotov15}.
The reference text in the training datasets is tokenized into phonetic inputs; the waveform is directly synthesized with a given x-vector $\z$ and GST $\g$ computed using the original audio data.
When the recordings are noisy, the x-vectors and GSTs also become noisy.
Therefore, we use only the clean speech for the speaker/style conditions, i.e., worn binaural recordings for CHiME-6 and the original WSJ dataset for CHiME-4.
%Because the x-vectors and GSTs of Kinect recordings are noisy, we used only worn binaural recordings for CHiME-6 setup.
Instead of shuffling speaker conditioning information, as in \cite{watanabe2020chime}, we perturb synthesized speech with various RIRs generated by a room simulator and additive noise for the CHiME-6 setup.
%As in \cite{watanabe2020chime}, the GMM-HMM model was reconstructed using an augmented dataset, and cleanup was applied to the TTS augmentation set based on the decoded scores.
%In CHiME-4, we used only the original WSJ dataset for the augmentation.

\subsection{Cycle-GAN data augmentation}
\label{ssec:cgan}
Cycle-GAN learns two mapping functions between two domains, given the sample sets of each domain \cite{zhu2017unpaired}.
%It is used in whisper speech recognition to augment data with synthesized whisper speech mapped from normal utterances \cite{gudepu2020whisper} and is also used in other speech studies \cite{hosseini2018multi, nidadavolu2019, bao2019cyclegan}.
It is used in speech domain adaptation by mapping between male and female speeches \cite{hosseini2018multi}, and is also used in other speech studies \cite{kaneko2018cyclegan, nidadavolu2019}. %, bao2019cyclegan}.
According to \cite{zorila2020toshiba}, GSS-based speech enhancement applied to the training Kinect data improves the ASR performance significantly because it aligns training more to the testing environment.
However, GSS-based speech enhancement can only be applied to multichannel data, which does not include TTS-synthesized data.
Therefore, instead, % of applying GSS to the training data, 
% we use Cycle-GAN to map the training data to the GSS-applied testing domain speech for CHiME-6 Challenge data.
% In CHiME-4 Challenge, we map the original WSJ data to the real recordings for the development set.
we use Cycle-GAN to map the clean training data to the speech-enhanced noisy testing speeches.
An advantage of using Cycle-GAN is that these two domain datasets do not need to be paired datasets.
In real situations, the paired data does not always exist; paired binaural data is a peculiar case for CHiME-6 Challenge.
In addition, the TTS synthesized speech is not strictly paired data to the target speech because the duration may vary or synthesis error may occur, which increases the difficulty of regression-based training.

Let the clean speech or TTS synthesized speech be $S$, and the target noisy observations be $X$.
Thus, Cycle-GAN trains two mapping functions $G:S\rightarrow X$ and $F:X\rightarrow S$ with two adversarial discriminators $D_S$ and $D_{X}$, where $D_S$ distinguishes between clean speech $S$ and translated speech $F(X)$ and so forth.
We further extend it to the multi-discriminator Cycle-GAN by following \cite{hosseini2018multi}.
The spectra $\S$ and $\X$ are computed from $\s$ and $\x$ with short-time Fourier transform (STFT) before divided into $m$ and $n$ frequency bands.
Subsequently, $D_S^{f_i}$ and $D_{X}^{f_i}$ are applied to each subband $f_i$ as the discriminators.

Generally, adversarial loss is computed as follows.
\begin{align}
    \mathcal{L}_{\mathrm{gan}}(G,D_{X}^{f_i\in n},S,X) =& \mathbb{E}\left[\sum_i^n\log D_{X}^{f_i}(\X)\right] \nonumber \\ +& \mathbb{E}\left[\sum_{i}^{n}\log(1-D_{X}^{f_i}(G(\S)))\right] 
\end{align}
In addition, cycle consistency loss is defined as 
\begin{align}
    \mathcal{L}_{\mathrm{cyc}}(G,F) =& \mathbb{E}\left[||F(G(\S))-\S||_1\right] \nonumber\\ &+ \mathbb{E}\left[||G(F(\X))-\X||_1\right],
\end{align}
where $||\cdot||_1$ is L1 loss.
Then, the total loss is optimized in combination with a tunable parameter $\lambda$.
\begin{align}
    \mathcal{L}(G,F,D_{S}^{f_i\in m},D_{X}^{f_i\in n}) =& \mathcal{L}_{\mathrm{gan}}(G,D_{X}^{f_i\in n},S,X) \nonumber\\ &+ \mathcal{L}_{\mathrm{gan}}(F,D_S^{f_i\in m},X,S) \nonumber \\ &+ \lambda \mathcal{L}_{\mathrm{cyc}}(G,F) \label{eq:cganloss}
\end{align}
Thus, by using only $G$ and the reference label $\y$, we obtain paired data by applying inverse-STFT as follows.
\begin{align}
    (\hat{\x}^{\mathsf{cgan}},\y) \leftarrow \hat{\x}^{\mathsf{cgan}} = \mathrm{ISTFT}(G(\S)) \label{eq:cgan}
\end{align}

\subsection{Pseudo-label augmentation}
Although pseudo-label augmentation is widely used in semi-supervised learning \cite{kemp1999unsupervised,%parthasarathi2019lessons,
chen2020semi%,xu2020iterative
}, we believe it is also effective for supervised setups where the labels are noisy because of conversational speech containing many ambiguities and disfluency.
The framework can also be considered as knowledge distillation \cite{kim16}, except that we can use the same model size for both teachers and students.
Beacuse of low resource data, the reference labels are biased; thus pseudo labels generated by statistical models such as neural networks would help interpolate its distribution.

Pseudo labels are estimated by a pretrained ASR model, as 
\begin{align}
    (\x,\hat{\y}^{\mathsf{pl}}) \leftarrow \hat{\y}^{\mathsf{pl}} = \argmax_{\y} p_{\mathrm{pre}}(\y|\x). \label{eq:pl}
\end{align}
Although a rigorous knowledge distillation for RNN-T was proposed in \cite{panchapagesan2020efficient}, we use the pseudo-label sequence directly as a reference sequence because it is efficient and sufficiently effective.

Because distant-talk transcription is a challenging task, the pseudo label may contain a noticeable number of errors.
Therefore, the pseudo labels were filtered based on the character error rates (CERs).
Given the CER threshold $\delta$, pseudo labels not greater than $\delta$ are kept in $\hat{Y}^{\mathsf{pl}}_{\delta}$.
We investigate the effectiveness of the filtering in Section~ \ref{sssec:filtering}.

\subsection{Combination of the augmentation methods}
The aforementioned augmentation methods can be used in combinations.
%The loss computation for RNN-T is defined using a forward--backward probability $p(\y|\x)$ \cite{graves13rnnt} as follows.
%\begin{align}
%    \mathcal{L}_{\mathrm{rnnt}} &= - \sum_{\x,\y\in(X, Y)} %\log p(\y|\x)
%\end{align}
With all the augmented paired data from (\ref{eq:tts}), (\ref{eq:cgan}), and (\ref{eq:pl}), the combined loss is computed as 
\begin{align}
    \mathcal{L} =& -\hspace{-0.1cm}\sum_{\x,\y\in (X,Y)} \hspace{-0.1cm} \log p(\y|\x) - \hspace{-0.3cm}\sum_{\ \ \  \hat{\s}^{\mathsf{tts}},\y \in (\hat{S}^{\mathsf{tts}},Y)} \hspace{-0.3cm}\log p(\y|\hat{\s}^{\mathsf{tts}}) \nonumber \\
    & -\hspace{-0.4cm}
    \sum_{\ \ \ \hat{\x}^{\mathsf{cgan}},\y \in (\hat{X}^{\mathsf{cgan}},Y)} \hspace{-0.3cm}\log p(\y|\hat{\x}^{\mathsf{cgan}}) -\hspace{-0.4cm}
    \sum_{\ \ \ \x,\hat{\y}^{\mathsf{pl}}\in (X,\hat{Y}^{\mathsf{pl}}_{\delta})} \hspace{-0.3cm}\log p(\hat{\y}^{\mathsf{pl}}|\x).
\end{align}
In the case of RNN-T, $p(\y|\x)$ is a forward--backward probability \cite{graves13rnnt}.
We further investigate every possible combinations in the following section.

\section{Experiments}
\subsection{CHiME-6 Challenge}
\subsubsection{Experimental setup}
We carried out experiments on the CHiME-6 Challenge dataset.
It consisted of 44 h of worn binaural data as well as multiple Kinect recordings.
The dataset was prepared following the baseline system in \cite{watanabe2020chime}, which included the perturbation using RIRs generated by a room simulator and the speed perturbation with a factor of \{0.9, 1.0, 1.1\}.
%For the training set, a subset of 400k Kinect utterances was combined with worn binaural data perturbed with various RIRs generated by a room simulator and additive noise.
%Subsequently, GMM-HMM was trained for data cleaning before speed perturbation was applied with a warping factor of \{0.9, 1.0, 1.1\}.
The baseline training set was in total 1400 h.
The development and evaluation sets were two dinner party sessions respectively.
We applied GSS-based enhancement \cite{boeddeker2018front}, following the baseline setup, as in \cite{watanabe2020chime}.

We extracted log-Mel filterbanks % as well as x-vectors 
using the Kaldi toolkit.
The output units were characters (26 alphabets and 21 auxiliary symbols).
%Following \cite{andrusenko2020towards}, RNN-T was trained with adadelta optimization \cite{zeiler2012adadelta} as a baseline.
%The model comprised four visual geometry group convolutional neural networks (VGG-CNNs) followed by a six-layer bi-LSTM with 512 units, and a RNN-T of a two-layer LSTM with 256 units.
%All layers were trained with a dropout rate of 0.4.
As a baseline, RNN-T was trained following  \cite{andrusenko2020towards}.
SpecAugment \cite{park19} was applied to the training set.
All models were trained with 10 epochs, and last 5 models were averaged for regularization.
External word-level LSTM language model (LM) was also trained with CHiME-6 training text corpus.
The model consisted of a one-layer uni-directional LSTM with 500 units.
When the LM was used, decoding was done with the shallow fusion with a weight of 0.1, and a beam size of 10.
The training was carried out using ESPNet \cite{watanabeespnet}.

Transformer TTS was trained with the Librispeech corpus \cite{panayotov15} for augmentation conditioned on an x-vector predicted by a TDNN trained with Voxceleb \cite{nagrani2017voxceleb} and VoxCeleb2 \cite{chung2018voxceleb2}.
The reference text was tokenized into pronunciation using the CMU dictionary and fed into the TTS module, which directly predicted the speech signal sequence.
The synthesized speech was further perturbed with RIRs as in Section~\ref{ssec:tts}.
Using both synthesized speech and original CHiME-6 training data, the GMM-HMM was trained, which was used for cleanup based on the decoded scores.
The aforementioned speed perturbation was then applied, which ended up augmenting the data to 2070 h.

Two Cycle-GANs were trained with two pairs of domains: binaural worn recordings and GSS applied enhanced training Kinect set, and TTS synthesized speech and the GSS training set.
Residual networks were trained following \cite{zhu2017unpaired}, except the fact that input and output features were spectrograms normalized with the mean and variance.
%Only generators $G$ in Section~\ref{ssec:cgan} were used to map from the worn data and TTS data to the GSS-enhanced domain, followed by speed perturbation.
We set $m=n=3$ in (\ref{eq:cganloss}). %Section~\ref{ssec:cgan}.
The Cycle-GANs followed by the speed perturbation added merely 123 h to the original set, but when it was combined with TTS augmentation the data increased up to 2366 h.

\begin{comment}
\begin{table}[t]
  \caption{Total data sizes with augmentation.}
  \label{tab:datasize}
  %\vspace{1mm}
  \vspace{-0.3cm}
  \centering
%  \begin{tabular}{l|cc|cc}
  \scalebox{0.9}{
  \begin{tabular}{l|c}
    \hline
     & hrs  \\
    \hline\hline
    baseline & 1400 \\
    \hline
    + TTS Aug. & 2070 \\
    + Cycle-GAN Aug. & 1523 \\
    + Pseudo Label Aug. & 2376 \\
    \hline
    + TTS + Cycle-GAN Aug. & 2366 \\
    + Cycle-GAN + Pseudo Label Aug. & 2548 \\
    + TTS + Pseudo Label Aug. & 3146 \\
    + All combined & 3512 \\
    \hline
    \end{tabular}
  }
  \vspace{-0.4cm}
\end{table}
\end{comment}

The baseline RNN-T model was also used for generating pseudo labels.
All generated labels were evaluated once, and values that were not greater than $\delta=50$ of the CER were used for augmentation.
%, because it was the best in the development set in our preliminary experiment.
This nearly doubled the dataset from 1400 h to 2630 h.
When all the proposed augmentation methods were combined, the dataset reached 3988 h in total.
The data sizes are summarized in the last column in Table~\ref{tab:results}.

\begin{table}[t]
  \caption{WERs of RNN-T E2E models in the CHiME-6 task}
  \label{tab:results}
  %\vspace{1mm}
  \vspace{-0.3cm}
  \centering
%  \begin{tabular}{l|cc|cc}
  \scalebox{0.87}{
  \hspace{-0.5cm}
  \begin{tabular}{l|cc|cc|c}
    \hline
    & \multicolumn{2}{c|}{w/o LM} & \multicolumn{2}{c|}{w/ LM} &\\
     & dev  & eval & dev  & eval & hrs\\
    \hline\hline
    TDNN-F hybrid \cite{watanabe2020chime} &&& 51.8 & 51.3 &1400\\
    % TDNN-F hybrid (reprod.) \cite{watanabe2020chime} &&& 50.2 & 50.9 &1400\\
    Joint CTC/Attention E2E \cite{Dalmia2018} &&& 82.1 & 71.8& \\
    CNN Multi-ch. E2E \cite{yalta2019cnn} &&& 80.7 & -& \\
    \hline
    RNN-T E2E (reprod.) \cite{andrusenko2020towards} & 59.5 & 58.2 &58.7 & 56.7 &1400\\ % w/o average (61.4/61.1/61.2/59.6) 0.3lm 58.9/56.1
    + SpecAugment (reprod.) \cite{andrusenko2020towards} & 55.8 & 55.7 & 55.4 & 54.4& \\ % w/o average (56.4/55.5/55.5/54.5) 0.3lm 55.6/54.4
     %\ \ \ \ {\it large model}& {\it 57.8} & {\it 56.3} & {\it 56.4} & {\it 55.1} \\ %0.3lm 56.2/54.8
    \hline
    ++ TTS Aug. & 54.4 & 53.9 & 53.6 & 52.8 &2070\\ % w/o average (54.5/55.0//53.9) 0.3lm 54.3/53.8
    %++ Cycle-GAN Aug. & 54.4& 53.9 & 53.9 & 53.3 &1523\\%0.3lm 55.0/55.1
    ++ Cycle-GAN Aug. & 54.2& 52.6 & 53.8 & 52.0 &1523\\ % cgantr
    ++ Label Aug. & 55.1 & 54.4 & 54.6 & 53.0 &2630\\
    % ++ 20CER Label Aug. & 55.5 & 54.9 & 55.0 & 53.5 &2376\\% w/o average (55.2/54.6//53.0) 0.3lm 54.8/53.0
    %\ \ \ \ {\it + Incremental training} &55.5 & 54.9&55.1&53.0\\ 
    \hline
    ++ TTS + Cycle-GAN Aug. & 51.5 & 50.7 &51.1 & 50.5 &2366 \\ % cgantr
    %++ TTS + Cycle-GAN Aug. & 52.4 & 53.8 &51.9 & 53.5 &2366 \\%55.1(53.3) & 57.2(57.4) & & (56.3)\\ 0.3lm 52.6/55.6
    %++ Cycle-GAN + Label Aug. & 53.8 & 52.7 &53.1 & 52.1 &2834\\
    ++ Cycle-GAN + Label Aug. & 53.7 & 52.1 &53.0 & 51.7 &2834\\ % cgantr
    % ++ Cycle-GAN + 20CER Label Aug. & 54.2 & 52.8 &54.0 & 52.4 &2548\\%(54.9) & (54.4) && (53.0)\\ 
    ++ TTS + Label Aug. & 53.1  & 53.3 & 52.1 & 51.7 &3630\\
    % ++ TTS + 20CER Label Aug. & 52.8  & 53.4 & 52.6 & 52.1 &3146\\ % 0.3lm 52.9/52.6
    %++ All combined & {\bf 51.7} & {\bf 51.5} &{\bf 51.0} & {\bf 50.8} &3988\\
    ++ All combined & {\bf 50.9} & {\bf 50.5} &{\bf 50.4} & {\bf 50.1} &3988\\ % cgantr
    % ++ 20 CER All combined & {\bf 52.0} & {\bf 51.5} &{\bf 51.7} & {\bf 51.0} &3512\\ %0.3lm 52.3/52.7
     %\ \ \ \ {\it + Incremental training } & & &&\\
     %\ \ \ \ {\it large model} & Apr 4 & &&\\
     \hline
     %\multicolumn{5}{l}{with {\it gss\_train } } \\
     \hline
     + SpecAugment + {\it gss\_train} \cite{andrusenko2020towards} & 53.6 & 51.4 & 53.7 & 51.0 & 1437 \\ % 0.3lm 54.2/52.4
     %++ All combined & {\bf 50.2} & {\bf 49.4} & {\bf 49.8} & {\bf 49.5} & 4025 \\
     ++ All combined & {\bf 49.6} & {\bf 48.7} & {\bf 49.5} & {\bf 48.6} & 4025 \\ % cgantr
     %\hline
     %JHU & & & 43.25 &\\
     \hline
  \end{tabular}
  }
  \vspace{-0.4cm}
\end{table}

\subsubsection{ASR results in the CHiME-6 task}
We first evaluate the effectiveness of each proposed augmentation method.
For comparison, other E2E ASR architectures are listed as well as the baseline RNN-T model in Table~\ref{tab:results}.
% The baseline RNN-T with SpecAugment achieved performance comparable to that of the TDNN-F hybrid system.
Although, Andrusenko {\it et al.} reported in \cite{andrusenko2020towards} that LM fusion degraded its performance, we observed that the external LM improved the WERs consistently. %, except with the {\it train\_gss} augmentation.

With each proposed augmentation on top of the SpecAugment, the WER was respectively reduced comparing to only the SpecAugment (54.4\% in eval set).
%Interestingly, using Cycle-GAN, which increased only one hour of data, improved ASR performance.
Pseudo label augmentation significantly improved the WER (53.0\%), which indicated that it is not only effective for unsupervised learning but also supervised learning with noisy label scenarios.
Among three methods, Cycle-GAN augmentation archived the best result (52.0\%).
 
Further, we investigated combinations of the three augmentation methods.
The combination of TTS and Cycle-GAN augmentation significantly dropped the error rates, and when all three were combined, we achieved the best WER (50.1\%).
Our proposed augmentation methods reduced the WER on eval set by 4.3\%, which was more significant than the WER reduction of the SpecAugment (2.3\%).
Improvements were also seen when we applied our proposed methods to TDNN-F hybrid model\footnote{We obtained marginal improvements in TTS/Label augmentation (1.0\%/1.6\% WER reductions) while Cycle-GAN degraded by 1.1\%.  Combinations were not always promising in TDNN-F.  We leave it as our future investigation.}.

According to \cite{zorila2020toshiba}, train set enhanced by GSS improved the recognition accuracy.
Therefore, we also included {\it train\_gss} set in training and the word error rate (WER) of dev set was reduced by 2.2\% absolute without LM, which almost matched the report in \cite{andrusenko2020towards} (2.4\% reduction).
On top of that, the combination of our proposed augmentation methods successfully reduced the WERs further.

\subsubsection{Effectiveness of pseudo label filtering}
\label{sssec:filtering}
We also investigated the effectiveness of filtering for pseudo-label augmentation.
The threshold $\delta$ was sampled from \{20, 50, 70, $\infty$\}, and the results were compared with the one without augmentation.
The results are shown in Table~\ref{tab:label_filter}.
% When we used all the pseudo labels, the WER of dev set increased (55.6\% $\rightarrow$ 55.8\%), which indicated that filtering played an important role.
$\delta=50$ achieves the best performance in the dev set; $\delta=70$ was the best in the eval set.
% Even with $\delta=20$, the data are still augmented by a factor of approximately 1.7.

\begin{comment}
\subsubsection{Effectiveness on TDNN-F hybrid models}
We also investigated the effectiveness of proposed augmentation methods on the baseline TDNN-F hybrid architecture.
The results are listed in Table~\ref{tab:hybrid_result}.
TTS augmentation and pseudo-label augmentation slightly improved its performance, while Cycle-GAN augmentation degraded.
These results indicate that the proposed methods are effective particularly for E2E ASR models.
\end{comment}

\subsection{CHiME-4 Challenge}
\subsubsection{Experimental setupt}
We conducted experiments on the CHiME-4 Challenge to confirm that our proposed augmentation methods were also effective for other datasets and model architectures.
As the baseline, the Conformer \cite{gulati2020} model was trained following \cite{guo2020recent}.
SpecAugment was also applied to the task.
Each training set was trained for 100 epochs and 10 models with the best validation accuracies were averaged for regularization.
We also trained an external word-level one-layer LSTM LM with 1000 units using CHiME-4 training text corpus.
The LM was fused with a weight of 1.0 and the beam size was fixed to 6.

For augmentation, we used the same Librispeech TTS model to generate synthesized speeches as in the CHiME-6 task.  
Cycle-GAN for CHiME-4 Challenge was trained using the original WSJ dataset and the real isolated recording for the development set.  
Pseudo labels were generated using the baseline Conformer model.
We found that approximately 90\% of the pseudo labels did not contain errors, i.e., $\mathrm{CER}=0$.
Therefore, we excluded these labels and used only the remains. We applied filtering using $\delta=2$, because our preliminary experiments provided the best results in the development set.
The duration of training data is summarized in the last column in Table~\ref{tab:chime4_results}. 

\subsubsection{ASR results in the CHiME-4 task}
First, we evaluated individual augmentation methods.
The results are summarized in Table~\ref{tab:chime4_results}.
All the proposed methods improved WERs in both simulation and real dataset.
Pseudo-label augmentation was not as significant as the other methods, which indicated that the reference label contained few errors and was not as biased as the CHiME-6 dataset.

Subsequently, therefore, we combined TTS and Cycle-GAN augmentations, which provided the best WER in real development set.
%When we combined all the proposed methods, we achieved 12.7\% in simulation set and 12.2\% in real set.
Finally, the combination of all augmentations provided a slight improvement in the simulation evaluation set.

\section{Conclusion}
We investigated three data augmentation methods for E2E ASR in the distant-talk scenarios.
TTS-synthesized speech was used for data augmentation which was further perturbed by applying RIRs and additive noise.
Cycle-GAN was also used to augment domain-matched speech by mapping from cleand recordings or TTS speech to the target noisy recordings.
Finally, pseudo-label augmentation was proposed for the supervised scenarios to smoothen the label distribution, where the labels were also noisy.
Experiments on the CHiME-6 and CHiME-4 tasks indicated that our proposed augmentation methods were effective in the RNN-T/Conformer E2E ASR models.
Further, combinations of those improved its performance, particularly when all the methods were combined in the CHiME-6 setup.
% We achieved 3.6\% of WER reduction with data augmentation upon SpecAugment.
%We achieved WER reduction as significant as that of the SpecAugment with the combination.

Future work includes on-the-fly augmentation of TTS and pseudo-labels, by introducing consistency losses, and shuffling speaker information. %, and virtual adversarial training.
%Particularly, 
Pseudo labels can also be used in a KL-divergence style as in knowledge distillation studies.
%We are also interested in applying these augmentation methods to other tasks such as AMI corpus.
%We are also interested in applying these augmentation methods to the conventional hybrid ASR models.

\begin{table}[t]
  \caption{Pseudo label augmentation with various filtering criteria.}
  \label{tab:label_filter}
  %\vspace{1mm}
  \vspace{-0.3cm}
  \centering
%  \begin{tabular}{l|cc|cc}
  \scalebox{0.9}{
  \begin{tabular}{l|cc|c}
    \hline
    Filtering criteria & dev  & eval & hrs \\
    \hline\hline
    No pseudo label & 55.4 & 54.4 & 1400 \\
    $\leq$20 CER & 55.0 & 53.5 & 2376 \\
    $\leq$50 CER & {\bf 54.6} & 53.0 & 2630 \\
    $\leq$70 CER & 54.9 & {\bf 52.7} & 2691 \\
    All pseudo label & 55.2 & 53.5 & 2799\\ % w/o average, w/o rnnlm (55.5/55.5) 0.3lm 55.8/53.9
    \hline
  \end{tabular}
  }
  \vspace{-0.2cm}
\end{table}

\begin{comment}
\begin{table}[t]
  \caption{WERs of TDNN-F hybrid models in the CHiME6 task.}
  \label{tab:hybrid_result}
  %\vspace{1mm}
  \vspace{-0.3cm}
  \centering
%  \begin{tabular}{l|cc|cc}
  \scalebox{0.9}{
  \begin{tabular}{l|cc|c}
    \hline
     & dev  & eval & hrs \\
    \hline\hline
    TDNN-F hybrid (reprod.) \cite{watanabe2020chime} & 50.2 & 50.9 & 1400 \\
    \hline
    + TTS Aug. & 49.8 & {\bf 49.7} & 2070 \\
    + Cycle-GAN Aug. & 53.8 & 51.9 & 1523 \\
    + Label Aug. &{\bf 48.8} & 49.2& 2630 \\
    + All combined & 50.0 & {\bf 49.7} & 3988\\
    \hline
  \end{tabular}
  }
  \vspace{-0.2cm}
\end{table}
\end{comment}

\begin{table}[t]
  \caption{WERs of Conformer E2E models for the CHiME-4 task}
  \label{tab:chime4_results}
  %\vspace{1mm}
  \vspace{-0.3cm}
  \centering
%  \begin{tabular}{l|cc|cc}
  \scalebox{0.9}{
  \hskip-0cm\begin{tabular}{l|cc|cc|c}
    \hline
    &\multicolumn{2}{c|}{dev} & \multicolumn{2}{c|}{eval} &\\
     & sim & real & sim & real & hrs \\
    \hline\hline
    Conformer \cite{guo2020recent} & 13.6&12.0 & 21.5&22.1 & 190 \\ 
    % 9.6,8.2,16.0,14.7
    % (10.6)(9.4)(17.6)(16.2)
    + SpecAugment \cite{guo2020recent} & 11.0&9.3 & 17.4&16.2 &  \\
    % 7.4,6.7,12.7,11.9
    % (9.8)(8.2)(15.4)(14.3)
    \hline
    ++ TTS Aug. & 10.5&8.7 & 17.3&15.8 & 253 \\
    % 6.9,5.8,11.5,10.7
    % (8.6)(7.2)(13.4)(12.5)
    ++ Cycle-GAN Aug. & \textbf{10.2}&8.5 & 17.1&\textbf{15.5} & 271 \\
    % 6.9,6.2,11.4,11.0
    % (9.1)(8.2)(13.1)(12.8)
    ++ Label Aug. & 10.7&9.1 & 17.2&16.4 & 202 \\
    % 7.3,6.3,12.2,11.5
    % (9.3)(7.7)(14.8)(13.8)
    \hline
    ++ TTS + Cycle-GAN Aug. & 10.4&\textbf{8.4} & 16.3&15.7 & 334 \\
    % 6.8,5.9,10.8,10.8
    % (8.7)(7.4)(12.4)(12.3)
    ++ All combined & 10.4&8.7 & \textbf{16.1}&15.7 & 346 \\
    % 7.0,5.9,10.9,10.8
    % (8.9)(7.7)(12.7)(12.2)
    %++ Label Aug. & 10.9&9.8 & 17.2&15.9 & 378 \\
    %++ 3\% Label Aug. & 9.5&8.2 & 14.7&13.8 & 197 \\
    %++ 5\% Label Aug. & 9.3&7.7 & 14.8&13.8 & 202 \\
    %++ 8\% Label Aug. & 9.5&8.2 & 14.7&13.7 & 208 \\
    \hline
  \end{tabular}
  }
  \vspace{-0.4cm}
\end{table}

\bibliographystyle{IEEEtran}

\bibliography{mybib}

\begin{thebibliography}{10}
\providecommand{\url}[1]{#1}
\csname url@samestyle\endcsname
\providecommand{\newblock}{\relax}
\providecommand{\bibinfo}[2]{#2}
\providecommand{\BIBentrySTDinterwordspacing}{\spaceskip=0pt\relax}
\providecommand{\BIBentryALTinterwordstretchfactor}{4}
\providecommand{\BIBentryALTinterwordspacing}{\spaceskip=\fontdimen2\font plus
\BIBentryALTinterwordstretchfactor\fontdimen3\font minus
  \fontdimen4\font\relax}
\providecommand{\BIBforeignlanguage}[2]{{%
\expandafter\ifx\csname l@#1\endcsname\relax
\typeout{** WARNING: IEEEtran.bst: No hyphenation pattern has been}%
\typeout{** loaded for the language `#1'. Using the pattern for}%
\typeout{** the default language instead.}%
\else
\language=\csname l@#1\endcsname
\fi
#2}}
\providecommand{\BIBdecl}{\relax}
\BIBdecl

\bibitem{graves06}
A.~Graves, S.~Fern{\'a}ndez, F.~Gomez, and J.~Schmidhuber, ``Connectionist
  temporal classification: labelling unsegmented sequence data with recurrent
  neural networks,'' in \emph{Proc. of 23rd International Conference on Machine
  Learning}, 2006, pp. 369--376.

\bibitem{chorowski15}
J.~K. Chorowski, D.~Bahdanau, D.~Serdyuk, K.~Cho, and Y.~Bengio,
  ``Attention-based models for speech recognition,'' in \emph{Proc. of NIPS},
  2015, pp. 577--585.

\bibitem{chan16}
W.~Chan, N.~Jaitly, Q.~Le, and O.~Vinyals, ``Listen, attend and spell: A neural
  network for large vocabulary conversational speech recognition,'' in
  \emph{Proc. of ICASSP}, 2016, pp. 4960--4964.

\bibitem{watanabe17}
S.~Watanabe, T.~Hori, S.~Kim, J.~R. Hershey, and T.~Hayashi, ``Hybrid
  {CTC}/attention architecture for end-to-end speech recognition,''
  \emph{Journal of Selected Topics in Signal Processing}, vol.~11, no.~8, pp.
  1240--1253, 2017.

\bibitem{graves13rnnt}
A.~Graves, A.-R. Mohamed, and G.~Hinton, ``Speech recognition with deep
  recurrent neural networks,'' in \emph{Proc. of ICASSP}, 2013, pp. 6645--6649.

\bibitem{karita19}
S.~Karita, N.~Chen, T.~Hayashi, T.~Hori, H.~Inaguma, Z.~Jiang, M.~Someki,
  N.~E.~Y. Soplin, R.~Yamamoto, X.~Wang \emph{et~al.}, ``A comparative study on
  transformer vs {RNN} in speech applications,'' in \emph{Proc. of ASRU
  Workshop}, 2019, pp. 449--456.

\bibitem{gulati2020}
A.~Gulati, J.~Qin, C.-C. Chiu, N.~Parmar, Y.~Zhang, J.~Yu, W.~Han, S.~Wang,
  Z.~Zhang, Y.~Wu \emph{et~al.}, ``Conformer: Convolution-augmented transformer
  for speech recognition,'' in \emph{Proc. of Interspeech}, 2020, pp.
  5036--5040.

\bibitem{ko2015audio}
T.~Ko, V.~Peddinti, D.~Povey, and S.~Khudanpur, ``Audio augmentation for speech
  recognition,'' in \emph{Sixteenth Annual Conference of the International
  Speech Communication Association}, 2015.

\bibitem{jaitly2013vocal}
N.~Jaitly and G.~E. Hinton, ``Vocal tract length perturbation ({VTLP}) improves
  speech recognition,'' in \emph{Proc. ICML Workshop on Deep Learning for
  Audio, Speech and Language}, vol. 117, 2013.

\bibitem{park19}
D.~S. Park, W.~Chan, Y.~Zhang, C.-C. Chiu, B.~Zoph, E.~D. Cubuk, and Q.~V. Le,
  ``{SpecAugment}: A simple data augmentation method for automatic speech
  recognition,'' in \emph{Proc. of Interspeech}, 2019.

\bibitem{hayashi2018back}
T.~Hayashi, S.~Watanabe, Y.~Zhang, T.~Toda, T.~Hori, R.~Astudillo, and
  K.~Takeda, ``Back-translation-style data augmentation for end-to-end {ASR},''
  in \emph{2018 IEEE Spoken Language Technology Workshop (SLT)}, 2018, pp.
  426--433.

\bibitem{tjandra2019end}
A.~Tjandra, S.~Sakti, and S.~Nakamura, ``End-to-end feedback loss in speech
  chain framework via straight-through estimator,'' in \emph{Proc. of ICASSP},
  2019, pp. 6281--6285.

\bibitem{wang2020improving}
G.~Wang, A.~Rosenberg, Z.~Chen, Y.~Zhang, B.~Ramabhadran, Y.~Wu, and P.~Moreno,
  ``Improving speech recognition using consistent predictions on synthesized
  speech,'' in \emph{Proc. of ICASSP}, 2020, pp. 7029--7033.

\bibitem{kemp1999unsupervised}
T.~Kemp and A.~Waibel, ``Unsupervised training of a speech recognizer: Recent
  experiments,'' in \emph{Sixth European Conference on Speech Communication and
  Technology}, 1999.

\bibitem{chen2020semi}
Y.~Chen, W.~Wang, and C.~Wang, ``Semi-supervised {ASR} by end-to-end
  self-training,'' in \emph{Proc. of Interspeech}, 2020, pp. 2787--2791.

\bibitem{du2020ustc}
J.~Du, Y.-H. Tu, L.~Sun, L.~Chai, X.~Tang, M.-K. He, F.~Ma, J.~Pan, J.-Q. Gao,
  D.~Liu \emph{et~al.}, ``The {USTC-NELSLIP} systems for {CHiME-6} challenge,''
  in \emph{Proc. of CHiME-6 Workshop}, 2020.

\bibitem{chen2020ioa}
H.~Chen, P.~Zhang, Q.~Shi, and Z.~Liu, ``The {IOA} systems for {CHiME-6}
  challenge,'' in \emph{Proc. of CHiME-6 Workshop}, 2020.

\bibitem{Dalmia2018}
S.~Dalmia, S.~Kim, and F.~Metze, ``{Situation informed end-to-end {ASR} for
  noisy environments},'' in \emph{Proc. of CHiME-5 Workshop}, 2018.

\bibitem{yalta2019cnn}
N.~Yalta, S.~Watanabe, T.~Hori, K.~Nakadai, and T.~Ogata, ``{CNN}-based
  multichannel end-to-end speech recognition for everyday home environments,''
  in \emph{2019 27th European Signal Processing Conference (EUSIPCO)}, 2019,
  pp. 1--5.

\bibitem{andrusenko2020towards}
A.~Andrusenko, A.~Laptev, and I.~Medennikov, ``Towards a competitive end-to-end
  speech recognition for {CHiME}-6 dinner party transcription,'' in \emph{Proc.
  of Interspeech}, 2020, pp. 319--323.

\bibitem{zhu2017unpaired}
J.-Y. Zhu, T.~Park, P.~Isola, and A.~A. Efros, ``Unpaired image-to-image
  translation using cycle-consistent adversarial networks,'' in
  \emph{Proceedings of the IEEE International Conference on Computer Vision},
  2017, pp. 2223--2232.

\bibitem{watanabe2020chime}
S.~Watanabe, M.~Mandel, J.~Barker, and E.~Vincent, ``{CHiME-6} challenge:
  Tackling multispeaker speech recognition for unsegmented recordings,'' in
  \emph{Proc. of CHiME-6 Workshop}, 2020.

\bibitem{boeddeker2018front}
C.~Boeddeker, J.~Heitkaemper, J.~Schmalenstroeer, L.~Drude, J.~Heymann, and
  R.~Haeb-Umbach, ``Front-end processing for the {CHiME-5} dinner party
  scenario,'' in \emph{Proc. of CHiME-5 Workshop}, 2018.

\bibitem{vincent2017analysis}
E.~Vincent, S.~Watanabe, A.~A. Nugraha, J.~Barker, and R.~Marxer, ``An analysis
  of environment, microphone and data simulation mismatches in robust speech
  recognition,'' \emph{Computer Speech \& Language}, vol.~46, pp. 535--557,
  2017.

\bibitem{li2019neural}
N.~Li, S.~Liu, Y.~Liu, S.~Zhao, and M.~Liu, ``Neural speech synthesis with
  transformer network,'' in \emph{Proceedings of the AAAI Conference on
  Artificial Intelligence}, vol.~33, no.~01, 2019, pp. 6706--6713.

\bibitem{snyder2018x}
D.~Snyder, D.~Garcia-Romero, G.~Sell, D.~Povey, and S.~Khudanpur, ``X-vectors:
  Robust dnn embeddings for speaker recognition,'' in \emph{Proc. of
  ICASSP}.\hskip 1em plus 0.5em minus 0.4em\relax IEEE, 2018, pp. 5329--5333.

\bibitem{wang2018style}
Y.~Wang, D.~Stanton, Y.~Zhang, R.-S. Ryan, E.~Battenberg, J.~Shor, Y.~Xiao,
  Y.~Jia, F.~Ren, and R.~A. Saurous, ``Style tokens: Unsupervised style
  modeling, control and transfer in end-to-end speech synthesis,'' in
  \emph{International Conference on Machine Learning}.\hskip 1em plus 0.5em
  minus 0.4em\relax PMLR, 2018, pp. 5180--5189.

\bibitem{panayotov15}
V.~Panayotov, G.~Chen, D.~Povey, and S.~Khudanpur, ``{LibriSpeech}: an {ASR}
  corpus based on public domain audio books,'' in \emph{Proc. of ICASSP}, 2015,
  pp. 5206--5210.

\bibitem{hosseini2018multi}
E.~Hosseini-Asl, Y.~Zhou, C.~Xiong, and R.~Socher, ``A multi-discriminator
  {CycleGAN} for unsupervised non-parallel speech domain adaptation,'' in
  \emph{Proc. Interspeech}, 2018, pp. 3758--3762.

\bibitem{kaneko2018cyclegan}
T.~Kaneko and H.~Kameoka, ``Cyclegan-{VC}: Non-parallel voice conversion using
  cycle-consistent adversarial networks,'' in \emph{2018 26th European Signal
  Processing Conference (EUSIPCO)}, 2018, pp. 2100--2104.

\bibitem{nidadavolu2019}
P.~S. {Nidadavolu}, J.~{Villalba}, and N.~{Dehak}, ``Cycle-{GANs} for domain
  adaptation of acoustic features for speaker recognition,'' in \emph{Proc. of
  ICASSP}, 2019, pp. 6206--6210.

\bibitem{zorila2020toshiba}
C.~Zorila, M.~Li, D.~Hayakawa, M.~Liu, N.~Ding, and R.~Doddipatla,
  ``Toshiba’s speech recognition system for the chime 2020 challenge,'' in
  \emph{Proc. of CHiME-6 Workshop}, 2020.

\bibitem{kim16}
Y.~Kim and A.~M. Rush, ``Sequence-level knowledge distillation,'' in
  \emph{Proceedings of the 2016 Conference on Empirical Methods in Natural
  Language Processing}, 2016, pp. 1317--1327.

\bibitem{panchapagesan2020efficient}
S.~Panchapagesan, D.~S. Park, C.-C. Chiu, Y.~Shangguan, Q.~Liang, and
  A.~Gruenstein, ``Efficient knowledge distillation for {RNN}-transducer
  models,'' \emph{arXiv preprint arXiv:2011.06110}, 2020.

\bibitem{watanabeespnet}
S.~Watanabe, T.~Hori, S.~Karita, T.~Hayashi, J.~Nishitoba, Y.~Unno, N.~E.~Y.
  Soplin, J.~Heymann, M.~Wiesner, N.~Chen \emph{et~al.}, ``{ESPnet}: End-to-end
  speech processing toolkit,'' in \emph{Proc. of Interspeech}, 2019, pp.
  2207--2211.

\bibitem{nagrani2017voxceleb}
A.~Nagrani, J.~S. Chung, and A.~Zisserman, ``Voxceleb: a large-scale speaker
  identification dataset,'' in \emph{Proc. of Interspeech}, 2017, pp.
  2616--2620.

\bibitem{chung2018voxceleb2}
J.~S. Chung, A.~Nagrani, and A.~Zisserman, ``Voxceleb2: Deep speaker
  recognition,'' in \emph{Proc. of Interspeech}, 2018, pp. 1086--1090.

\bibitem{guo2020recent}
P.~Guo, F.~Boyer, X.~Chang, T.~Hayashi, Y.~Higuchi, H.~Inaguma, N.~Kamo, C.~Li,
  D.~Garcia-Romero, J.~Shi \emph{et~al.}, ``Recent developments on espnet
  toolkit boosted by conformer,'' \emph{arXiv preprint arXiv:2010.13956}, 2020.

\end{thebibliography}

\end{document}